\documentclass[epj]{svjour}

\usepackage{graphics}
\usepackage{epsfig,color}
\begin{document}
\title{Triplet superconductivity vs.\ easy-plane ferromagnetism in a 
1D itinerant electron system with transverse spin anisotropy}
\titlerunning{Triplet superconductivity vs.\ easy-plane ferromagnetism
in a 1D itinerant electron system}
%
%
\author{C. Dziurzik, G.I. Japaridze \thanks{\emph{Permanent address:} 
Andronikashvili Institute of Physics, Georgian 
Academy, Tamarashvili 6, Tbilisi 380077, Georgia. Electronic address: 
japa@iphac.ge}, A. Schadschneider \and J. Zittartz} 
\authorrunning{C.\ Dziurzik et al.}
\offprints{A. Schadschneider (as@thp.uni-koeln.de)}       
%
\institute{Institut f\"ur Theoretische Physik, Universit\"at zu K\"oln, 
50937 K\"oln, Germany}
\date{Received: \today }

\abstract{In this paper we study the ground state phase diagram of a 
one-dimensional $t-U-J$ model, at half-filling. In the large-bandwidth 
limit and for ferromagnetic exchange with easy-plane anisotropy, a phase 
with gapless charge and massive spin excitations, characterized by the 
coexistence of triplet superconducting and spin density wave 
instabilities is realized in the ground state. With reduction 
of the bandwidth, a transition into an insulating phase showing properties 
of the spin-$\frac{1}{2}$ $XY$ model takes place.}   


\maketitle

\section{Introduction}\label{intro}
Superconductivity near a magnetic instability is a topic of increased
current interest in condensed matter physics. Magnetically mediated
Cooper pairing near the antiferromagnetic instability is widely
discussed in the context of superconductivity in copper-oxide systems
\cite{Chubukov1}.  Moreover, the discovery of Triplet
Superconductivity (TS) in $Sr_{2}RuO_{4}$ \cite{MacKenzieMaeno} and
the recent discovery of coexistence of the TS phase with
ferromagnetism in $UGe_{2}$ \cite{UGE2}, $URhGe$ \cite{URhGe} and
$ZrZn_2$ \cite{ZrZn2} has triggered an increased activity in studies of
correlated electron models showing close proximity of triplet
superconducting and ferromagnetically ordered phases
\cite{RS,MS,Sigrist,Kirpatrick1,SinghMazin,WalkerSamokhin,Chubukov2,Buzdin,Kirpatrick2}.

Another group of unconventional superconductors with close
proximity of magnetic and superconducting ordering belongs to the
$(TMTSF)_{2}X$ family of quasi-one-dimen\-sional conductors 
(Bechgaard salts) \cite{Jerome1,Ishiguro}. Growing experimental
evidence has been collected in the last few years, indicating that 
$(TMTSF)_{2}ClO_{4}$ and $(TMTSF)_{2}PF_{6}$ under
pressure are triplet superconductors \cite{Trexp8}.
Most interesting is the phase diagram of $(TMTSF)_{2}PF_{6}$
which shows a spin-Peierls (SP) phase in the ground state at
atmospheric pressure. Increasing pressure leads first to a transition
from the SP phase into a spin density wave (SDW) phase, and finally to the 
suppression of the SDW ground state in favor of superconductivity
\cite{Jerome2}. Recent detailed experimental studies of the phase diagram 
of the $(TMTSF)_{2}PF_{6}$ compound indicate the possibility of a 
{\em coexistence regime} between SDW and 
superconductivity \cite{Jerome3}. Although the very presence of a 
\emph{homogeneous} coexistence phase in the phase diagram was questioned 
in the more recent publication \cite{Kornilov}, the SDW-SC competition is 
common in organic materials \cite{Jerome4} and therefore models of 
correlated electrons exhibiting such phases are of great interest.

Various models of strongly correlated electrons showing close
proximity of magnetic (ferromagnetic) and superconducting (triplet
superconducting) phases have been subject of intensive research in
attempt to construct a theoretical model for new superconducting
materials. Usually these models are based on some extensions of
the Hubbard model. In particular, several extended versions of the
repulsive Hubbard model have been employed for a long time as
standard models for metal-insulator transitions, antiferromagnetism
and high-$T_{c}$ superconductivity \cite{Dagotto}. At the same time,
the Hubbard model in the case of sufficiently narrow band and/or low
doping is a standard model for metallic ferromagnetism of itinerant
electrons \cite{FerroHUbbard}.

Taking into account the experimentally observed easy-plane 
anisotropy of the spin exchange \cite{Mukuda} in some of these materials, 
Japaridze and M\"uller-Hartmann \cite{JM} proposed a rather simple extension 
of the Hubbard model with transverse ($XY$-type) anisotropy as a suitable 
approach to such systems with coexisting orders. Indeed, this model was shown 
to exhibit an extremely rich \emph{weak-coupling} phase diagram. In 
particular in the case of a half-filled band the weak-coupling ground state 
phase diagram consists of two insulating antiferromagnetic phases with 
easy-plane anisotropy and a spin gapful metallic phase with an identical 
decay of the \emph{triplet superconducting} and spin density wave (SDW$^{(z)}$)
instabilities. Strong evidence for the presence of an additional
transition into a \emph{ferromagnetic} $XY$ phase has also been given
\cite{JM}.

The model describes a system of itinerant electrons with transverse
spin-exchange interaction between electrons on nearest-neighbor sites.
The one-dimensional version of the Hamiltonian reads:
\begin{eqnarray}
  \label{tJmodel}
  {\cal H} & = &  -t\sum_{n,\alpha}(c^{\dagger}_{n,\alpha}
  c^{\vphantom{\dagger}}_{n+1,\alpha}
  + c^{\dagger}_{n+1,\alpha}c^{\vphantom{\dagger}}_{n,\alpha}) \nonumber\\
  &+& U \sum_{n}\rho_{\uparrow}(n)\rho_{\downarrow}(n) +
  \frac{J_{xy}}{2}\sum_{n}(S^{+}_{n}S^{-}_{n+1} + h.c.)\,.
\end{eqnarray}
Here $c^{\dagger}_{n,\alpha}$ ($c^{\vphantom{\dagger}}_{n,\alpha}$) is
the creation (annihilation) operator for an electron at site $n$ with spin
${\alpha}$, $\rho_{,\alpha}(n)=c^{\dagger}_{n,\alpha}
c^{\vphantom{\dagger}}_{n,\alpha}$, $\vec{S}(n)=\frac{1}{2}
c^{\dagger}_{n,\alpha}\vec{\sigma}^{\vphantom{\dagger}}_{\alpha\beta}
c^{\vphantom{\dagger}}_{n,\beta}$
where $\sigma^{{\it i}}$ (${\it i}=x,y,z$) are the Pauli matrices.
Below we restrict our consideration to the case of repulsive on-site
interaction $U\geq0$ while the sign of the exchange interaction is
arbitrary.

One can easily verify that besides the obvious $U(1)$ spin symmetry in the
half-filled case the model is characterized by a $SU(2)$ charge symmetry.
An electron-hole transformation for one spin component interchanges the charge
and spin degrees of freedom, and maps (\ref{tJmodel})
to the attractive Hubbard model with pair-hopping interaction \cite{PK,JK}. 

That the TS phase can be realized in 1D correlated electron systems is
well known from standard ``{\em g-ology}'' studies \cite{Solyom}.
The extended ($U$-$V$) Hubbard model with nearest-neighbor attraction
($V<0$) has been intensively studied to explain the competition between 
SDW and superconducting instabilities in $TMTSF$ compounds \cite{Fukuyama}. 
However, due to spin rotational invariance, in the extended Hubbard model 
the TS phase is realized only in the Luttinger liquid phase for $|U|<-2V$ 
\cite{Solyom,Emery,Voit}, where both charge and spin excitations are gapless. 
Singlet superconducting (SS) and TS correlations show identical power-law 
decay at large distances and the TS instability dominates only due to weak 
logarithmic corrections \cite{Voit}. On the other hand, in the spin gapped 
phase $U<2V$, the dynamical generation of a spin gap leads to 
the {\em complete suppression} of the TS and SDW instabilities. In marked 
contrast, the ferromagnetic transverse exchange between electrons on 
neighboring sites provides the possibility for realization of the SDW-TS 
phase in the case of gapped spin excitation spectrum \cite{JM}.

In this paper we study the model (\ref{tJmodel}) in the case of
a half-filled band using the DMRG techniques. We investigate the
excitation spectrum of the system as well as the behavior of various
correlation functions.  Our numerical results confirm the predictions
of a weak-coupling analysis.  In addition, we study in detail the
strong-coupling sector of the phase diagram, focusing our attention on
the ferromagnetic transition. With increasing transverse ferromagnetic
exchange this will reveal the possibility for
\emph{two different} scenarios of transition into the easy-plane $XY$
ferromagnetic phase: in the case of weak on-site repulsion ($U<U_{c}
\simeq 2t$) the first transition at $J_{xy}=J^{(c1)}_{xy}$ takes
place from a spin gapped metallic phase into the insulating SDW$^{(z)}$
phase with long-range order (LRO) and the latter becomes unstable towards 
the spin-flop transition
into the \emph{ferromagnetic} $XY$ phase at $J^{(c2)}_{xy}$. On
the other hand, in the case of strong on-site repulsion, the metallic
phase is absent and only the spin-flop transition from the SDW$^{(z)}$
phase into the \emph{ferromagnetic} $XY$ phase takes place at
$J^{(c2)}_{xy}$. The critical value of the exchange strongly
depends on the on-site repulsion. At $U=0$, $J^{(c1)}_{xy}=-3t$ and
$J^{(c2)}_{xy}=-4t$. In the case of weak $U \ll t$ these parameters are
renormalized linearly in $U$, $J^{(c1)}_{xy}=0$ at $U \simeq 2t$, while 
for $U \gg t$ we have $J^{(c2)}_{xy} \sim 1/U$.

The paper is organized as follows: in the next section the
weak-coupling continuum-limit version of the model is investigated. 
In Sect.~\ref{sec:NumericalResults} results of DMRG studies 
for chains up to $L=120$ sites are presented. 
Finally, Sect.~\ref{sec-conclusion} is devoted to a discussion and 
concluding remarks.

\section{\bf The continuum-limit theory}
\label{sec-continuum}
In this section we consider the low-energy effective field theory of the 
initial lattice model. Although this procedure has a long history 
and is reviewed in many places \cite{GNT}, for clarity we briefly sketch the 
most important points and fix our notation and conventions. Considering the  
weak-coupling limit, $|U|, |J_{xy}| \ll t$ we linearize the spectrum and 
pass to the continuum limit by substituting
\begin{equation}
  c_{n,\alpha} \rightarrow {\it i}^{n}R_{\alpha}(x) 
  + (-{\it i})^{n}L_{\alpha}(x)
  \label{RLdef}
\end{equation}
$x=na_{0}$, where $a_{0}$ is the lattice spacing, and $R_{\alpha}(x)$ and 
$L_{\alpha}(x)$ describe the R({\em ight}) and L({\em eft}) excitations with 
dispersion relations $E = \pm v_{F}p$. These fields are assumed to be 
smooth on the scale of the lattice spacing and can be bosonized in a 
standard way \cite{GNT}
\begin{eqnarray}\label{Bosonization1}
  R_{\alpha}(x)& = & \frac{1}{\sqrt{2\pi a_{0}}} 
  e^{{\it i}\sqrt{4\pi}\Phi_{R,\alpha}}(x)\,,\\
  L_{\alpha}(x)& = & \frac{1}{\sqrt{2\pi a_{0}}} 
  e^{-{\it i}\sqrt{4\pi}\Phi_{L,\alpha}}(x)
\end{eqnarray}
where $\Phi_{R(L),\alpha}(x)$ are the Right (Left) moving Bose fields, 
carrying spin $\alpha $.  Next we define
\begin{equation}
  \phi_{\alpha}=\phi_{L,\alpha}+\phi_{R,\alpha} \,, \qquad
  \theta_{\alpha}=\phi_{L,\alpha}-\phi_{R,\alpha}.
\end{equation}
and introduce linear combination 
\begin{eqnarray}
  \varphi_{c}&=&
  (\phi_{\uparrow}+ \phi_{\downarrow})/\sqrt{2}, \qquad \vartheta_{c}=
  (\theta_{\uparrow} +\theta_{\downarrow})/\sqrt{2}\,, \\
  \varphi_{s}&=&
  (\phi_{\uparrow}-\phi_{\downarrow})/\sqrt{2}, \qquad \vartheta_{s}=
  (\theta_{\uparrow} -\theta_{\downarrow})/\sqrt{2}  
  \label{BOSONS} 
\end{eqnarray}
to describe the \emph{charge} $(c)$  and \emph{spin} $(s)$ degrees of freedom, 
respectively. Then the Hamiltonian density of the bosonized model is given by
\begin{eqnarray}
  {\cal H} &=& {\cal H}_{c}+{\cal H}_{s},
  \nonumber\\
  {\cal H}_{c} &=& v_{c} \int dx\Big\{{1\over 2}[(\partial_{x}\varphi_{c})^2
  +(\partial_x \vartheta_{c})^2]
  \nonumber\\
  &+& \frac{M^{0}_{c}}{a_0^2}\cos(\sqrt{8\pi K^{0}_{c}}\varphi_{c})\Big\},
  \label{SGc}\\
  {\cal H}_{s} &=& v_{s}\int dx\Big\{{1\over 2}[(\partial_{x}\vartheta_{s})^2
  +(\partial_x\varphi_{s})^2]
  \nonumber\\
  &+& \frac{M^{0}_{s}}{a_0^2}\cos(\sqrt{8\pi K^{0}_{s}}\varphi_{s})\Big\}
  \label{SGs}.
\end{eqnarray}
Here we have defined
\begin{eqnarray}
  2\left(K^{0}_{c}-1\right) &=&
  g_{c} = - {1 \over \pi \tilde{v}_{F}}\left(U+J_{\perp}\right),
  \label{Kc}\\
  2\pi M^{0}_{c}& = &  g_{u} = - {1 \over \pi
    \tilde{v}_{F}}\left(U+J_{\perp}\right),
  \label{Mc}\\
  2\left(K^{0}_{s}-1\right)& = & g_{s} =
  {1 \over \pi \tilde{v}_{F}}\left(U+J_{\perp}\right),\label{Ks}\\
  2\pi M^{0}_{s}& = & g_{\perp}  =
  {1 \over \pi \tilde{v}_{F}}\left(U-J_{\perp}\right),
  \label{Ms}\\
  v_{c (s)}& = & {\tilde{v}_{F} \over K^{0}_{c(s)}}, \qquad
  \tilde{v}_{F}=2t\left(1 + {J_{\perp}\over 2\pi t}\right)\ .
  \label{Fv}
\end{eqnarray}
This mapping of the lattice electron system model onto the quantum
theory of two independent quantum Bose fields described in terms of an
''effective'' sine-Gordon (SG) Hamiltonians (\ref{SGc}) and (\ref{SGs})
will allow to extract the ground state properties of the initial model
using the far-infrared properties of the quantum SG theory.

The infrared behavior of the SG Hamiltonian is described by the
corresponding pair of renormalization group (RG) equations for the
effective coupling constants $K_{c(s)}(l)$ and $M_{c(s)}(l)$
\begin{eqnarray}
  \label{RGeq}
  \frac{dM_{c(s)}(l)}{dl} &=& -2\left(K_{c(s)}(l)-1\right)
  M_{c(s)}(l)
  \nonumber\\
  \frac{dK_{c(s)}(l)}{dl} &=& - \frac{1}{2}M_{c(s)}^{2}(l)
\end{eqnarray}
where $l=\ln(a_{0})$ and the bare values of the coupling constants are
$K_{c(s)}(l=0) \equiv K^{0}_{c(s)}$ and $M_{c(s)}(l=0)\equiv M_{c(s)}^{0}$.
The pair of RG equations (\ref{RGeq}) describes the Kosterlitz-Thouless
transition \cite{KT}.

The flow lines lie on the hyperbola
\begin{equation}
  \label{hyperbola}
  4\left(K_{c(s)} - 1\right)^{2}- M_{c(s)}^{2} = \mu^{2} =
  4(K^{0}_{c(s)}-1)^{2}-(M_{c(s)}^{0})^{2}
\end{equation}
and exhibit two different regimes depending on the relation between
the bare coupling constants \cite{Wiegmann}.

{\em Weak-coupling regime.} \ \ For $2(K_{c(s)}-1)\geq
\left|M^{0}_{c(s)}\right|$ we are
in the weak-coupling regime: the effective mass ${\cal M}_{c(s)} \to 0$. The
low energy (large distance) behavior of the corresponding gapless mode
is described by a free scalar field.

The vacuum averages of exponentials of the corresponding fields show
a power-law decay at large distances ($\eta\equiv c,s$)
\begin{eqnarray}
  \langle 
  e^{i\sqrt{2\pi K_{\eta}}\varphi_{\eta}(x)} 
  e^{-i \sqrt{2\pi K_{\eta}}\varphi_{\eta}(x')}
  \rangle 
  &\sim& \left| x - x' \right|^{- K_{\eta}},
  \label{freecorrelations1}\\
  \langle 
  e^{i \sqrt{2\pi/K_{\eta}}\vartheta_{\eta}(x)} 
  e^{-i\sqrt{2\pi/K_{\eta}}\vartheta_{\eta}(x')}
  \rangle
  &\sim& \left| x - x' \right|^{-1 /K_{\eta}},
  \label{freecorrelations2}
\end{eqnarray}
and the only parameter controlling the infrared behavior in the gapless
regime is the fixed-point value of the effective coupling constants
$K^{\ast}_{c(s)} = K_{c(s)}(l=\infty) $ determined from the
Eq.~(\ref{hyperbola}). Note that in the
$SU(2)$ symmetric case $\mu=0$ and $K^{\ast}_{c(s)}=1$.

{\em Strong coupling regime.} \ \ For $2(K^{0}_{c(s)}-1) <
\left|M^{0}_{c(s)}\right|$ the
system scales to strong coupling: depending on the sign of the bare
mass $M^{0}_{c(s)}$, the renormalized mass ${\cal M}_{c(s)}$ is driven to
$\pm\infty$,
signaling a crossover to one of two strong coupling regimes with a
dynamical generation of a commensurability gap in the excitation
spectrum. The flow of $\left|{\cal M}_{c(s)}\right|$ to large values indicates
that the ${\cal M}_{c(s)}\mbox{cos}\sqrt{8\pi K}\phi $ term in the
sine-Gordon model dominates the long-distance properties of the
system. Depending on the sign of the mass term, the field $\varphi$ gets
ordered with the expectation values \cite{ME}
\begin{equation}
  \label{orderfields}
  \langle\varphi_{c(s)}\rangle =\left\{ \begin{array}{l@{\quad}}
      \sqrt{\pi/8K_{c(s)}}
      \hskip0.5cm(M^{0}_{c(s)}>0) \\
      0 \hskip2.05cm (M^{0}_{c(s)}<0)\end{array}\right. \, .
\end{equation}

Using the initial values of the coupling constants, given in
(\ref{Kc})-(\ref{Ms}), we obtain that flow trajectories in the charge sector 
(due to the $SU(2)$-charge symmetry) are along the separatrix $g_{c}= g_{u}$. 
Therefore, at 
\begin{equation}
  \label{CSsc}
  U+J_{xy}>0.
\end{equation}
there is a gap in the charge excitation spectrum ($\Delta_{c} \neq 0$) and the 
charge field $\varphi_{c}$ is ordered  with the vacuum expectation value 
\begin{equation}
  \label{PHIcs}
  \langle  \varphi_{c} \rangle  = 0, 
\end{equation}
while at $U+J_{xy}<0$ the charge sector is gapless 
and the fixed-point value of the parameter $K^{\ast}_{c}$ is one.

The $U(1)$ symmetry of the spin channel ensures more alternatives. 
Depending on the relation between the bare coupling constants there are 
two different strong-coupling sectors in the spin channel. For 
\begin{equation}
  \label{SS1sc}
  U < \min\{0,J_{xy}\}
\end{equation}
the spin channel is massive ($\Delta_{s} \neq 0$) and the field $\varphi_{s}$ 
gets ordered with the vacuum expectation value 
\begin{equation}
  \label{PHI1ss}
  \langle  \varphi_{s} \rangle  = 0, 
\end{equation}
while for 
\begin{equation}
  \label{Ss1cc}
  J_{xy}< \min\{0,U\}
\end{equation}
the spin channel is massive ($\Delta_{s} \neq 0$), with vacuum 
expectation value
\begin{equation}
  \label{PHI2ss}
  \langle  \varphi_{s} \rangle  = \sqrt{\frac{\pi}{8K_{s}}}.
\end{equation}

In all other cases the excitation spectrum in the corresponding channel is 
gapless. The low-energy behavior of the system is controlled by the 
fixed-point value of the Luttinger-liquid parameter 
$K^{\ast}_{s}=1+\frac{1}{2}g^{\ast}_{s}$. In the particular case of 
vanishing on-site interaction ($U=0$) and antiferromagnetic exchange 
($J_{xy}>0$) one has to use a second order RG analysis to define 
accurately the fixed point value of the parameter $K_{s}$ (for details, 
see Ref.~\cite{JM}).

\subsection{Order parameters}
To clarify the symmetry properties of the ground states of the system in 
different sectors we consider the following set of order parameters: 
\begin{itemize}
\item[1)] the on-site density operator 
  $\rho(n)=\rho_{\uparrow}(n)+\rho_{\downarrow}(n)$
\end{itemize}
\begin{eqnarray}
  \label{Density}
  :\hat{\rho}(n): &=& \sum_{\alpha}(c^{\dagger}_{n,\alpha}
  c^{\vphantom{\dagger}}_{n,\alpha} - 1)   \simeq   
  \sqrt{\frac{2K_{c}}{\pi}}\partial_{x}\varphi_{c}
  \nonumber\\
  +&(-1)^{n}&
  \sin(\sqrt{2\pi K_c}\varphi_{c}) 
  \cos( \sqrt{2\pi K_{s}}\varphi_{s}) \, ,
\end{eqnarray}  
\begin{itemize}
\item[2)] the on-site spin-density
\end{itemize}
\begin{eqnarray}
  \label{Sz}
  S_{z}(n)&\simeq&
  \sqrt{\frac{K_{s}}{2\pi}}\partial_{x}\varphi_{s}\nonumber\\
  +&(-1)^{n}&
  \cos(\sqrt{2\pi K_c}\varphi_{c}) \sin\left( \sqrt{2\pi K_{s}}\varphi_{s}
  \right)\,,\\
  S_{x}(n)& \simeq  & { {\it i} \over \pi a_{0}}
  \cos(\sqrt{2\pi K_{s}}\varphi_{s})\sin\left(\sqrt{\frac{2\pi}{K_{s}}}
\vartheta_{s}\right)
  \nonumber\\
  +&(-1)^{n}&
  \cos(\sqrt{2\pi K_c}\varphi_{c}) \sin\left( \sqrt{\frac{2\pi}{K_{s}}}
  \vartheta_{s}\right)\,,\\
  S_{y}(n)&\simeq& { {-\it i} \over \pi a_{0}}
  \cos(\sqrt{2\pi K_{s}}\varphi_{s})\cos\left(\sqrt{\frac{2\pi}{K_{s}}}
\vartheta_{s}\right)
  \nonumber\\
  +&(-1)^{n}&
  \cos(\sqrt{2\pi K_c}\varphi_{c})\sin\left(\sqrt{\frac{2\pi}{K_{s}}}
  \vartheta_{s}\right)
\end{eqnarray}  
and in addition we use superconducting order parameters corresponding to  
\begin{itemize}
\item[3a)] the on-site singlet
\end{itemize}
\begin{eqnarray}
  \label{OnSiteSinglet}
  {\cal O}_{S}^{\dagger}(n) &=& 
  c^{\dagger}_{n,\uparrow}c^{\dagger}_{n,\downarrow} \nonumber\\
  & \sim & \cos(\sqrt{2 \pi K_{s}}\varphi_{s}) 
  \exp\left(i \sqrt{\frac{2\pi}{K_{c}}}\vartheta_{c}\right) \nonumber\\
  &-&(-1)^{n} \sin(\sqrt{2\pi K_c}\varphi_{c}) 
  \exp\left({\it i} \sqrt{\frac{2\pi}{K_{c}}}\vartheta_{c}\right) 
\end{eqnarray}
\begin{itemize}
\item[3b)] the extended singlet
\end{itemize}
\begin{eqnarray}
  \label{ExtendedSinglet}
  {\cal O}_{ES}^{\dagger}(n) &=&\frac{1}{\sqrt{2}}
  \left(c_{n,\uparrow}^{\dagger}c_{n+1,\downarrow}^{\dagger} -
    c_{n,\downarrow}^{\dagger}c_{n+1,\uparrow}^{\dagger}\right)\nonumber\\
  & \sim &(-1)^{n}\cos(\sqrt{2\pi K_c}\varphi_{c})\exp\left({\it i} 
  \sqrt{\frac{2\pi}{K_{c}}}\vartheta_{c}\right) 
\end{eqnarray}
\begin{itemize}
\item[3c)] and the triplet pairing
\end{itemize}
\begin{eqnarray}\label{Triplet}
{\cal O}^\dagger_{TS^{0}}(n) &= & \frac{1}{\sqrt{2}}
\left(c_{n,\uparrow}^{\dagger}c_{n+1,\downarrow}^{\dagger} +
c_{n,\downarrow}^{\dagger}c_{n+1,\uparrow}^{\dagger}\right)\nonumber\\
&\sim& \sin(\sqrt{2 \pi K_{s}}\varphi_{s})
\exp\left({\it i} \sqrt{\frac{2\pi}{K_{c}}}\vartheta_{c}\right)
\end{eqnarray}
\begin{eqnarray}\label{Triplet2}
{\cal O}^\dagger_{TS^{\pm}}(n) &= & \frac{1}{\sqrt{2}}
\left(c_{n,\uparrow}^{\dagger}c_{n+1,\uparrow}^{\dagger} \pm
c_{n,\downarrow}^{\dagger}c_{n+1,\downarrow}^{\dagger}\right)\nonumber\\
&\sim&\exp\left({\it i} \sqrt{\frac{2\pi}{K_{c}}}\vartheta_{c}\right)
\left\{ \begin{array}{l@{\quad}}
\cos\big({\frac{2\pi}{K_{s}}}\vartheta_{s}\big) \\
\sin\big({\frac{2\pi}{K_{s}}}\vartheta_{s}\big)
\end{array}\right. \, .
\end{eqnarray}

Note that the smooth part in Eq.~(\ref{OnSiteSinglet}) corresponds
to the usual BCS-type pairing while the oscillating terms in
(\ref{OnSiteSinglet}) and (\ref{ExtendedSinglet}) describe the 
eta-pairing superconductivity \cite{Yang}.

\subsection{Phases}
\begin{figure}[here]
  \centering
  \input{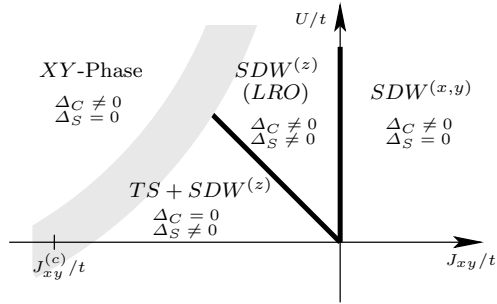}
  \caption{\label{fig:weak_phase}{The weak-coupling phase diagram of the 
      itinerant $XY$-model at half-filling. Solid lines indicate borders 
      between phases. The shaded region qualitatively marks the transition 
      into the $XY$ magnetic phase ($J_{xy}^{(c)}\approx-4t$).}}
\end{figure}

With the results of the previous section for the excitation spectrum 
and the behavior of the corresponding fields 
Eqs.~(\ref{freecorrelations1})--(\ref{orderfields}) we now analyze the 
\emph{weak-coupling} ground state phase diagram of the model 
(\ref{tJmodel}) (see Fig.~\ref{fig:weak_phase}).

Let us first consider the case $U=0$, where the weak-coupling analysis 
shows existence of two different phases: in the case of antiferromagnetic 
exchange, at $J_{xy}>0$, there is a gap in the charge excitation 
spectrum while the spin sector is gapless. Ordering of the field 
$\varphi_{c}$ with vacuum expectation value $\langle \varphi_{c} \rangle =0$ 
leads to a suppression of the CDW and \emph{superconducting} correlations. 
The SDW and \emph{Peierls} correlations show a power-law decay at large 
distances \cite{JM}. Due to the $U(1)$-spin symmetry, 
$K^{\ast}_{s}>1$ and the \emph{``in-plane''} correlations dominate in the 
ground state,
\begin{eqnarray}
  \label{SxSx}
  \langle S^{+}(r) S^{-}(0)\rangle \sim  
  r^{-K^{\ast}_{s}-1/K^{\ast}_{s}} + (-1)^{r} r^{-1/K^{\ast}_{s}}
\end{eqnarray}
while the longitudinal spin correlations 
\begin{eqnarray}
  \label{SzSz}
  \langle S^{z}(r) S^{z}(0)\rangle \simeq r^{-2} + (-1)^{r} r^{-K^{\ast}_{s}}
\end{eqnarray}
and \emph{Peierls} correlations decay faster. 

We now focus on the case of ferromagnetic exchange between spins. At $U=0$
and $J_{xy}<0$ there is a gap in the spin excitation spectrum while the 
charge excitation spectrum is gapless.  As common in the half-filled band 
case, the gapless charge excitation spectrum opens a possibility for the 
realization of a \emph{superconducting} instability in the system. Moreover, 
due to the $U(1)$-symmetry of the system, ordering of $\varphi_{s}$ with 
vacuum expectation value $\langle \varphi_{s} \rangle =\sqrt{\pi/8K_{s}}$ 
leads to a suppression of the CDW and singlet correlations as well as 
$S^z=\pm 1$ channels of the triplet pairing. However, the $S^z-S^z$ and 
triplet correlations in the $S^z=0$ channel show an identical power-law 
decay 
\begin{eqnarray}
  \label{SzSz2}
  \langle S^{z}(r) S^{z}(0)\rangle =
  \langle {\cal O}^{\dagger}_{TS}(r){\cal O}_{TS}(0)\rangle 
  \simeq  (-1)^{r} r^{-1} 
\end{eqnarray}
at large distances and are the \emph{dominating instabilities} in the system.

Let us now consider the effect of the on-site Coulomb repulsion.
At $J_{xy}>0$ the easy-plane antiferromagnetic phase remains unchanged
at $U>0$. However, at $J_{xy}<0$ the TS$+$SDW$^{(z)}$ phase is stable only
towards influence of a weak $U <-J_{xy}$ on-site coupling.
In the case of repulsive Hubbard interaction, 
at $U>-J_{xy}$ a charge gap opens. This regime corresponds to the 
appearance of a long-range ordered \emph{antiferromagnetic (N\'eel)} phase
\begin{equation}
  \label{correlSDWz}
  \langle S^z(r)S^z(0)\rangle \sim {\it constant}
\end{equation}
in the ground state. 

\subsection{The ferromagnetic transition}
Let us now discuss the ferromagnetic transition in the itinerant $XY$
model ($U=0$). The very presence of this transition can already been
seen within the weak-coupling studies, however detailed analysis of
the phase diagram close to transition is out of scope of the
continuum-limit approach.  As we obtained, at $J_{xy}<0$, $|J_{xy}|
\ll t$, the charge excitation spectrum is gapless and the spin
excitation spectrum is massive.  However, in the limit of strong
ferromagnetic exchange $|J_{xy}| \gg t$, the model is equivalent to
the $XY$ spin chain. Therefore, with increasing coupling one has to
expect a transition from the regime with massive spin and massless
charge excitation spectrum into a insulating magnetic phase with
gapless spin excitations.  On the other hand, in the case of antiferromagnetic
exchange $J_{xy} > 0$ the weak-coupling study shows a phase with
gapless spin, gapped charge and dominating easy-plane spin
correlations. One expects that this phase evolves smoothly to the
strong coupling limit.

The $J_{xy} \leftrightarrow -J_{xy}$ asymmetry is already seen on the
level of the Hartree regularization of the band-width cut-off parameter
$W=2\pi t$ as given by the Eqs.~(\ref{Fv})
\begin{equation}
  W_{\it eff}=2\pi\left(1+\frac{J_{xy}}{2\pi t}\right).
\end{equation}
The weak-coupling approaches fail when the effective dimensionless
coupling constant 
$|g_{\it i}| = \frac{J_{xy}}{2\pi t} =|g^{c}_{\it i}|\simeq 1$.  
This condition immediately gives $J^{(c)}_{xy}=-\pi t$. As
we show below, using the DMRG studies of chains up to $L=120$ sites,
indeed the transition into the ferromagnetic easy-plane ordering 
discussed above takes place at $J^{(c)}_{xy} \sim -4t$.

\section{Numerical results}\label{sec:NumericalResults}
We use the density-matrix renormalization-group (DMRG) method
\cite{White92,Peschel99} to study the ground-state properties of this
model.  Our calculations have been performed for open chains up to
$120$ sites using the \emph{infinite-size} version of the DMRG
routine. A comparison with the \emph{finite-size} algorithm, which
requires more CPU time and memory, does not give a substantial
improvement of the results.  For most of the numerical results
reported here we have kept $400$ states in each block, which produces
truncation errors smaller than $10^{-7}$.

In order to reduce edge effects we average correlation functions $C(|i-j|)$ 
over a number of pairs $(i,j)$ of lattice sites separated by the same 
displacement $r:=|i-j|$ \cite{NWS}. 
Typically we take nine pairs and for each value $r$ we place 
the pairs as close to the center of the chain as possible.

The asymptotic behavior of correlations (e.g. exponents)
has been determined by an appropriate fitting of the data \cite{cd_phd}.

\subsection{Excitation spectrum at $U=0$}
\begin{figure}[here]
  \resizebox{1.05\columnwidth}{!}
  {\includegraphics{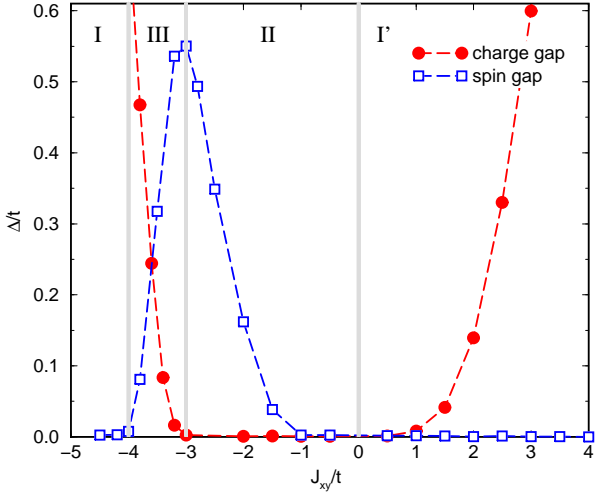}}
  \vspace{-0.8cm}
  \caption{Spin and charge excitation spectrum of the itinerant $XY$ model 
    at half-filling. Depending on the behavior of the gaps four sectors
    can be distinguished. The approximate boundaries 
    $J_{xy}^{(c1)}\approx-3t$ and $J_{xy}^{(c2)}\approx-4t$ are indicated by 
    grey lines.}
  \label{fig:spectrum_U0}      
\end{figure}

Let us start from the limiting case of the itinerant $XY$ model $(U=0)$ and 
analyze its excitation spectrum. The charge and spin gap for a half-filled 
$L$-site system are evaluated by 
\begin{eqnarray}
  \Delta_C(L) &=& \frac{1}{2}
  \left[E_0\left(\frac{L}{2}+1,\frac{L}{2}+1\right) + 
    E_0\left(\frac{L}{2}-1,\frac{L}{2}-1\right)\right.\nonumber\\
  && \hspace{5pt}\left.-2E_0\left(\frac{L}{2},\frac{L}{2}\right)\right]\,,\\
  \Delta_S(L) &=& E_0\left(\frac{L}{2}+1,\frac{L}{2}-1\right)-
  E_0\left(\frac{L}{2},\frac{L}{2}\right)\,,
  \label{eq:spin_charge_gap_def}
\end{eqnarray} 
respectively, where $E_0(N_{\uparrow},N_{\downarrow})$ is the ground-state 
energy for $N_{\uparrow}$ up-spin and $N_{\downarrow}$ down-spin electrons. 
The extrapolation for $L\rightarrow\infty$ is then performed by fitting a 
polynomial in $1/L$ to the calculated finite-chain results. 
Figure~\ref{fig:spectrum_U0} displays the extrapolated values as a function 
of $J_{xy}$.

We observe the following four sectors: at $J_{xy}>0$ the system is 
characterized by gapless spin and gapped charge excitation spectrum, while
the weak-coupling ferromagnetic sector exhibits gapless charge and gapful
spin degrees of freedom. 

Moreover, our numerical results show the presence of two new regions.
At $J_{xy}^{(c1)}\approx -3t$ a charge gap opens, while the spin gap
starts to decrease and finally closes at $J_{xy}^{(c2)}\approx -4t$.
This defines two new sectors: for $J_{xy}^{(c2)}<J_{xy}<J_{xy}^{(c1)}$
both the spin and charge sectors are gapped, while at
$J_{xy}<J_{xy}^{(c2)}$ the spin sectors become gapless. There are no
indications for further transitions in the system.  Note that similar
behavior of the gaps, with interchange of spin and charge degrees of
freedom, was first observed by Sikkema and Affleck in the Penson-Kolb
model \cite{SA}.

\subsection{Correlation functions at $U=0$}
To investigate the nature of ordering in the different phases we study the 
behavior of the correlation functions. 
In the sectors with gapless excitation spectrum and at half-filling
we expect the usual expression for correlation functions
\begin{equation}
  C(r)\equiv
  \langle{\cal O}^{\dagger}(r){\cal O}^{\vphantom{\dagger}}(0)\rangle \sim
  A_1r^{-\theta_1}+(-1)^rA_2r^{-\theta_2}
\end{equation}
consisting of a smooth part decaying with exponent $\theta_1$
and an oscillating part decaying with $\theta_2$.
In determining the asymptotics of correlation functions we focus on the 
dominating part given by $\theta=\min\{\theta_1,\theta_2\}$.

In the following we will present results for correlation functions in
different sectors of the phase diagram.

\subsubsection{Sectors \textrm{I} and \textrm{I'} 
  ($\Delta_C\neq 0$, $\Delta_S=0$): The $XY$-phases}
\begin{figure}[here]
  \resizebox{1.05\columnwidth}{!}
  {\includegraphics{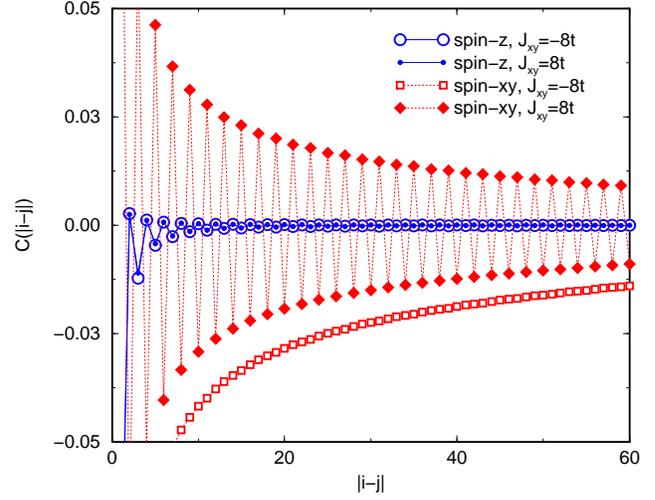}}
  \vspace{-0.8cm}
  \caption{The longitudinal (circle) and transverse 
    (square and diamond) spin-spin correlations in the case of strong 
    easy-plane ferromagnetic $J_{xy}=-8t$ (sector \textrm{I}) and 
    antiferromagnetic $J_{xy}=8t$ (sector \textrm{I'}) exchange.}
  \label{fig:spin_corr_xy_case} 
\end{figure}

In Fig.~\ref{fig:spin_corr_xy_case} we have plotted the longitudinal and 
transverse spin-spin correlations in the case of strong easy-plane exchange. 
Although the amplitudes of the transverse correlation functions are different,
the estimated exponents are similar. In the case of ferromagnetic exchange
we obtained $\theta\approx 0.57$, whereas for the antiferromagnetic exchange
we have $\theta\approx 0.61$. 
The results are in a good agreement with the
exact value $\theta=0.5$ obtained for the $XY$-model \cite{XY-Corr}. 
The longitudinal correlation functions decay faster. The calculated 
exponents $\theta\approx 1.79$ (for $J_{xy}=-8t$) and $\theta\approx 1.66$ 
(for $J_{xy}=8t$) are close to the exact XY-value $\theta= 2$. 

\begin{figure}[here]
  \resizebox{1.05\columnwidth}{!}{
    \includegraphics{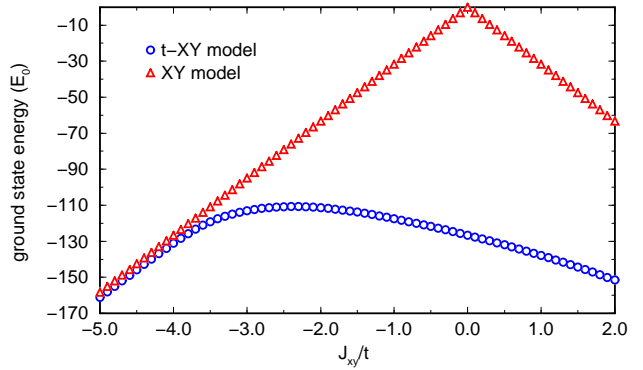}}
  \vspace{-2.4cm}
  \caption{Ground-state energy $E_0$ of the $XY$-model and the itinerant 
    $XY$-model as a function of coupling $J_{xy}/t$ for a half-filled $L=100$ 
    chain.}
  \label{fig:gnd_energy} 
\end{figure}

The asymmetry of this model is clearly seen in Fig.~\ref{fig:gnd_energy}, 
where the ground state energy as a function of $J_{xy}$ is presented.
As we observe from Fig.~\ref{fig:gnd_energy} in the case of ferromagnetic 
exchange the ground state energy of the itinerant model becomes very 
close to that of the spin-$\frac{1}{2}$ $XY$ chain.

\subsubsection{Sector \textrm{II} ($\Delta_C=0$, $\Delta_S\neq0$): 
  The TS+SDW$^{(z)}$ regime}

\begin{figure}[here]
  \resizebox{1.05\columnwidth}{!}
  {\includegraphics{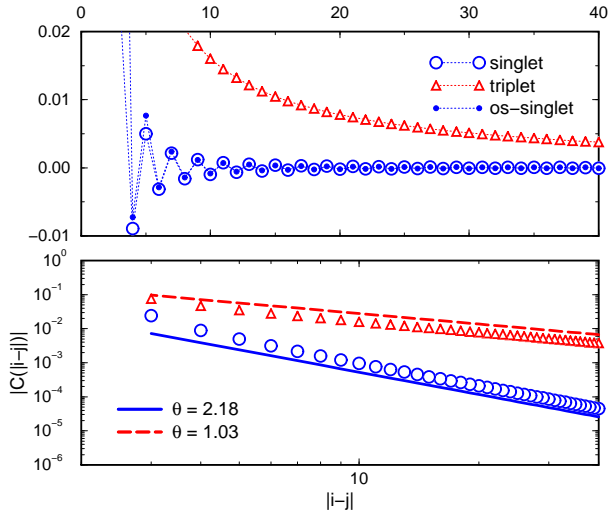}}
  \vspace{-0.8cm}
  \caption{Pair correlation functions in the case of ferromagnetic exchange
    $J_{xy}=-2t$ and $U=0$ (sector \textrm{II}). The lower figure shows the 
    algebraic decay of the triplet and singlet correlation, plotted on a 
    double logarithmic scale.} 
  \label{fig:pair_corr_Jxym2_U0} 
\end{figure}

Let us now focus on the case of ferromagnetic exchange $J_{xy}<0$ at $U=0$.
The bosonization results predict a suppression of the CDW and singlet
correlations, whereas SDW$^{(z)}$ and triplet correlators show identical 
power-law decay (cf. with Eq.~(\ref{SzSz2})). Furthermore, they are the 
dominating instabilities in this phase. Figure~\ref{fig:pair_corr_Jxym2_U0} 
displays DMRG results for the singlet- and triplet-pair correlation function.
One can clearly observe a strong triplet-pair correlation. 
Note that the on-site and extended singlet-pair correlations show an
almost identical behavior. This is expected from the bosonization
results (\ref{OnSiteSinglet}) and (\ref{ExtendedSinglet}) since
the smooth part of the on-site singlet correlations  (\ref{OnSiteSinglet})
does not contribute due to (\ref{PHI2ss}). In the double 
logarithmic plot (see lower figure) all correlation functions indicate a 
power-law decay with fast decaying singlet-pairing correlators
($\theta\approx 2.18$) and a slowly decaying triplet correlation function 
($\theta\approx 1.03$). The results are in a good agreement with those 
predicted by bosonization. 
\begin{figure}[here]
  \resizebox{1.05\columnwidth}{!}
  {\includegraphics{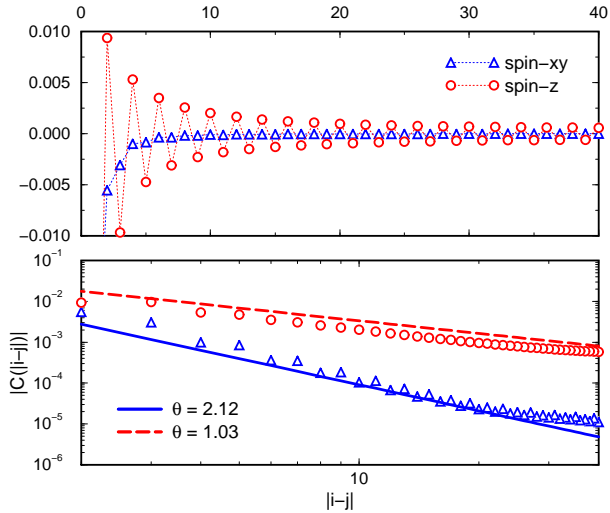}}
  \vspace{-0.8cm}
  \caption{DMRG results for the longitudinal and transversal spin correlation 
    function, plotted against the real space distance $|i-j|$ (upper figure) 
    at $J_{xy}=-2t$ (sector \textrm{II}). The exponents (lower figure) 
    were calculated using a suitable subset of the data
    to reduce finite size effects and 
    numerical inaccuracies at large distances.}  
  \label{fig:spin_corr_Jxym2_U0} 
\end{figure}

In Fig.~\ref{fig:spin_corr_Jxym2_U0} we show 
calculations for the longitudinal and transversal spin-spin correlation for 
ferromagnetic exchange ($J_{xy}=-2t$). We observe that the correlation 
functions exhibit an algebraic decay in which the transverse 
spin-spin correlation function decays faster. The calculated exponent of the 
longitudinal spin correlation function is, in agreement with bosonization 
results, close to that of the triplet-pairing correlations. 
Compared to the other results, the scaling behavior of the spin correlations
is not very good in this sector. However, we have verified that with 
increasing system size and number of states kept in the DMRG algorithm
the region with algebraic decay increases. Nevertheless, the numerical 
estimates for the exponents are less reliable than those for other
correlation functions.  

\begin{figure}[here]
  \resizebox{1.05\columnwidth}{!}
  {\includegraphics{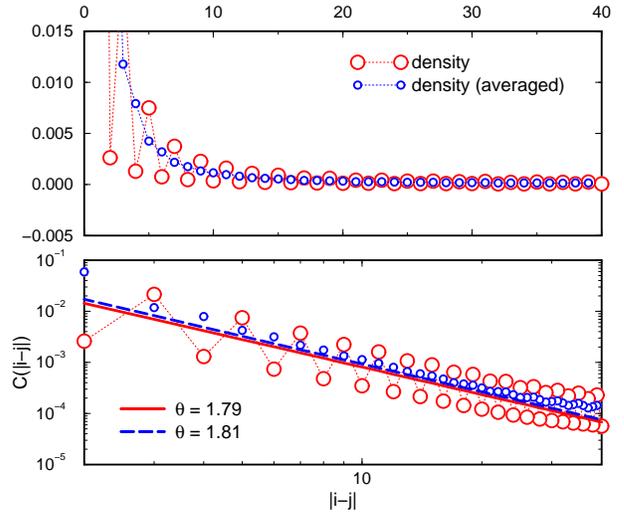}}
  \vspace{-0.8cm}
  \caption{DMRG results for the density-density correlation function at 
    $J_{xy}=-2t$ (sector \textrm{II}) including the average value of this 
    correlation which removes the even-odd-$r$ oscillations.}
  \label{fig:charge_corr_Jxym2_U0} 
\end{figure}

To complete the weak-coupling picture of sector II, we performed calculations 
for the density-density correlation. The results are shown in 
Fig.~\ref{fig:charge_corr_Jxym2_U0}.
Since in the double logarithmic plot we observe strong oscillations 
we additionally calculate the average value \cite{HHM95}
\begin{equation}
  \bar{C}(r) = \frac{1}{4}[C(r-1)+2C(r)+C(r+1)] 
\end{equation}
to smoothen the curve. As its clearly seen from the lower part of  
Fig.~\ref{fig:charge_corr_Jxym2_U0} the oscillations are removed, but
the estimated exponent remains almost unchanged. Thus the DMRG result
indicates a fast decay of density-density correlations, in agreement with 
the bosonization results. 

Therefore, we can conclude that coexisting triplet-pairing and 
antiferromagnetic SDW$^{(z)}$ ordering are the dominating instabilities in 
this sector.

\subsubsection{Sector \textrm{III} ($\Delta_C\neq 0$, $\Delta_S\neq 0$):
  The intermediate phase}
In this subsection we analyze the intermediate sector at $-4t\le J_{xy}\le -3t$
which is absent in the weak-coupling phase diagram 
(cf. with Fig.~\ref{fig:weak_phase}). 

We start to examine the asymptotic behavior of the superconducting 
and spin-spin correlations.
\begin{figure}[here]
  \resizebox{1.05\columnwidth}{!}
  {\includegraphics{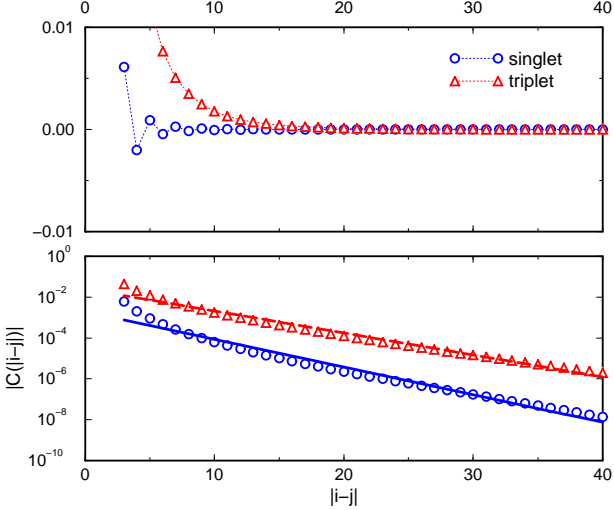}}
  \vspace{-0.8cm}
  \caption{Pair correlation functions at $J_{xy}=-3.5t$ and $U=0$ 
    (sector \textrm{III}). The lower part shows a double logarithmic plot 
    with linear fits.}
  \label{fig:pair_corr_Jxym3o5_U0} 
\end{figure}
\begin{figure}[here]
  \resizebox{1.05\columnwidth}{!}
  {\includegraphics{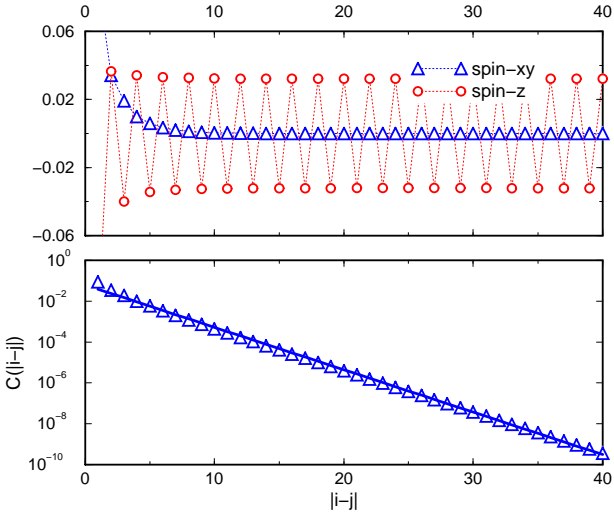}}
  \vspace{-0.8cm}
  \caption{Spin correlation functions at $J_{xy}=-3.5t$ and $U=0$ 
    (sector \textrm{III}). The lower part shows a double logarithmic plot 
    with a linear fit.}
  \label{fig:spin_corr_Jxym3o5_U0} 
\end{figure}
In Fig.~\ref{fig:pair_corr_Jxym3o5_U0} we present DMRG data for the pairing 
correlation functions. As is clearly seen from the figure, especially
from the logarithmic plot, the superconducting correlations decay
exponentially in agreement with the presence of a charge gap.

In Fig.~\ref{fig:spin_corr_Jxym3o5_U0} we plot the spin-spin correlation
functions. The logarithmic plot shows that the transverse spin
correlation functions decay exponentially. Contrary, the longitudinal
spin correlations show well-established long-range order.

The appearance of LRO is consistent with the $U(1)\otimes{Z}_2$ 
spin-symmetry of the present model (\ref{tJmodel}). 
The continuous $U(1)$ symmetry is generated by the operators $S^x$ and $S^y$,
while the discrete  ${Z}_2$ symmetry comes from the invariance
with respect to the $S^z\to -S^z$ transformation.
Since the SDW$^z$ ordering violates the discrete ${Z}_2$ and
translation symmetries, the true LRO state is not forbidden.

\subsection{Behavior for nonvanishing on-site interaction}
\subsubsection{Excitation spectrum at $U\neq 0$}
Let us now consider the effects of a repulsive Coulomb interaction
on the ground state phase diagram of the model. We start with the
excitation spectrum. 

From the bosonization results we know the general effect of the Coulomb 
repulsion on the phase diagram which displays itself in an enlargement 
of the charge gap sectors at the expense of the spin gap sector.
\begin{figure}[here]
  \resizebox{1.05\columnwidth}{!}
  {\includegraphics{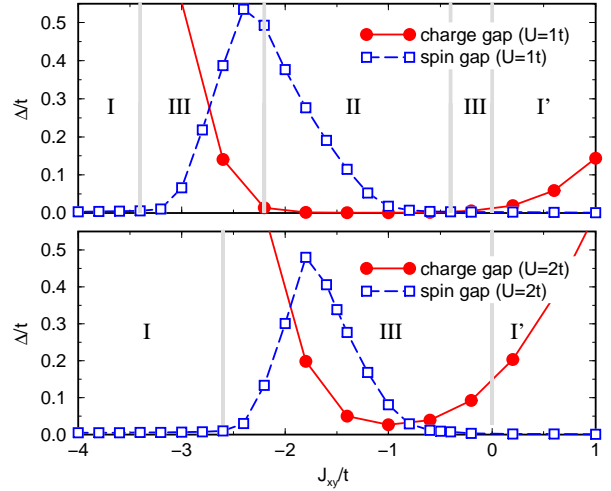}}
  \vspace{-0.8cm}
  \caption{Spin and charge excitation gaps of the itinerant $XY$ model 
    at $U=t$ (upper figure) and $U=2t$ (lower figure).}
  \label{fig:spectrum_U1and2}      
\end{figure}
Fig.~\ref{fig:spectrum_U1and2} shows charge and spin gaps
for $U=t$ and $U=2t$. One can clearly see that sectors I and I',
where we have a finite charge gap $\Delta_C>0$, are enlarged.
As a consequence the spin-gapped phase (sector II) 
becomes smaller with increasing $U$ and finally vanishes completely.
Already at $U=2t$ the charge gap is always finite.
Thus the main effect of the presence of Coulomb interactions
is the suppression of sector II, i.e. a reduction of the region
with dominating superconducting correlations. In analogy with the $U=0$ case
we conclude that the sectors with magnetic correlations become dominating.

\subsubsection{Correlation functions at $U\neq 0$}
In the following we analyze the effect of the Coulomb interactions on pair 
and spin correlation functions. We will focus on the behavior in sectors II 
and III where $\Delta_C=0$, $\Delta_S\neq 0$ and $\Delta_C\neq 0$, 
$\Delta_S\neq 0$, respectively.

\begin{figure}[here]
  \resizebox{1.05\columnwidth}{!}
  {\includegraphics{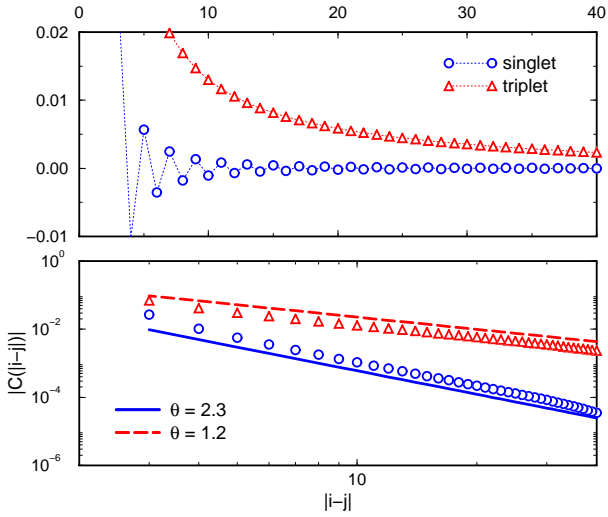}}
  \vspace{-0.8cm}
  \caption{Pair correlation functions in the ferromagnetic phase at 
    $J_{xy}=-1.5t$ and $U=t$ (sector \textrm{II}). The lower figure shows 
    the algebraic decay of the triplet and singlet correlation, plotted on 
    a double logarithmic scale.} 
  \label{fig:pair_corr_Jxym1o5_U1} 
\end{figure}

\begin{figure}[here]
  \resizebox{1.05\columnwidth}{!}
  {\includegraphics{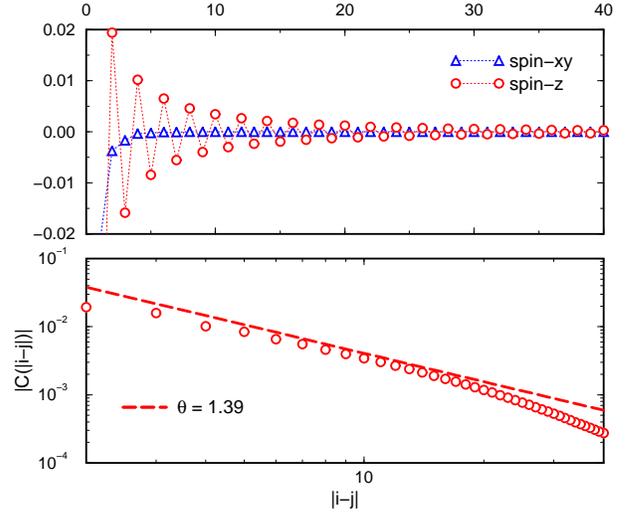}}
  \vspace{-0.8cm}
  \caption{Spin correlation functions at $J_{xy}=-1.5t$ and $U=t$ 
    (sector \textrm{II}). Lower figure is plotted on a double logarithmic 
    scale showing an algebraic decay for the longitudinal correlator and 
    exponentially decaying behavior for the transversal correlation function.}
  \label{fig:spin_corr_Jxym1o5_U1} 
\end{figure}

In the TS + SDW$^{(z)}$ phase we consider the coupling $J_{xy}=-1.5t$ at $U=t$
as a representative point. The phase is characterized by a spin gap of 
magnitude $\Delta_S\approx 0.13t$ and massless charge mode. The asymptotic 
behavior of the pair and spin
correlation functions is plotted in Fig.~\ref{fig:pair_corr_Jxym1o5_U1} and 
Fig.~\ref{fig:spin_corr_Jxym1o5_U1}, respectively. 
One can clearly see that the triplet-pairing and longitudinal spin-spin 
correlations represent the dominating instabilities in the system. 
Unfortunately the accuracy of the numerics is not sufficient in this case 
to verify that the exponents are still exactly identical. Instead, we find
$\theta\approx 1.2$ (triplet pairing) and $\theta\approx 1.39$ (longitudinal
spin). 

In the SDW$^{(z)}$ phase with LRO we compute the correlation functions 
at $U=2t$ and $J_{xy}=-2t$. The presence of a charge gap 
$\Delta_C\approx 0.38t$ leads now to an exponential decay of 
superconducting correlations. 
\begin{figure}[here]
  \resizebox{1.05\columnwidth}{!}
  {\includegraphics{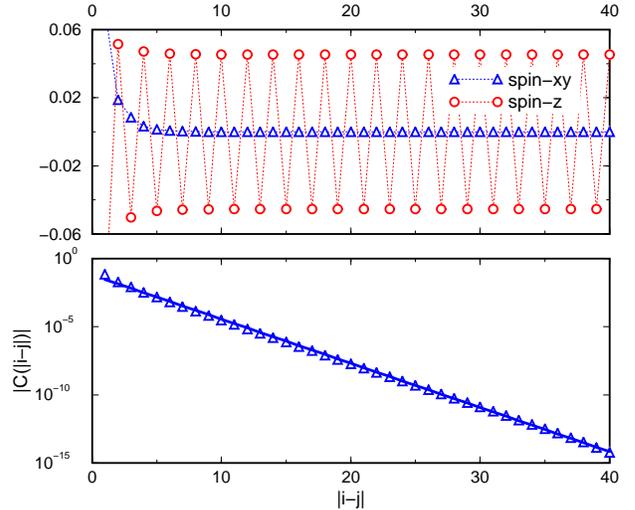}}
  \vspace{-0.8cm}
  \caption{Spin correlation functions at $J_{xy}=-2t$ and $U=2t$ 
    (sector \textrm{III}). Upper figure indicates LRO fluctuation for the 
    longitudinal spin correlation function.}
  \label{fig:spin_corr_Jxym2_U2} 
\end{figure}

On the other hand, as is clearly seen from Fig.~\ref{fig:spin_corr_Jxym2_U2},
the longitudinal spin-spin correlations show a true LRO while
the transverse spin correlations decay exponentially.

\begin{figure}[here]
  \resizebox{1.05\columnwidth}{!}
  {\includegraphics{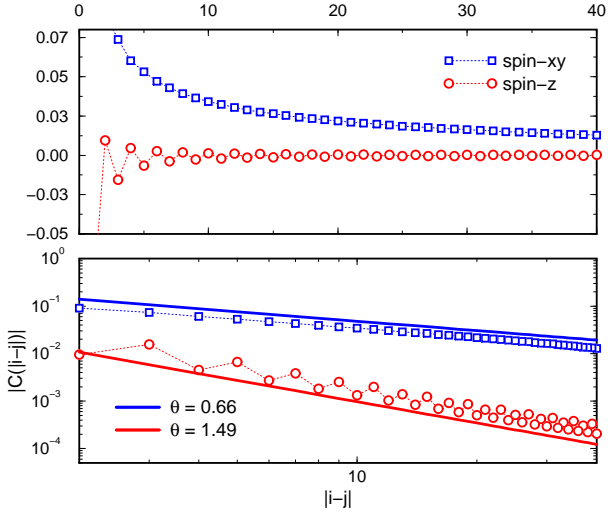}}
  \vspace{-0.8cm}
  \caption{Spin correlation functions at $J_{xy}=-4t$ and $U=2t$ 
    (sector \textrm{I}).}
  \label{fig:spin_corr_Jxym4_U2} 
\end{figure}

In ferromagnetic phase we use as a representative point $U=2t$ and 
$J_{xy}=-4t$. As one can observe from Fig.~\ref{fig:spin_corr_Jxym4_U2} the 
longitudinal spin-spin correlations are exponentially suppressed, while the 
decay of the transverse ferromagnetic spin correlations is almost identical 
to that of the standard $XY$-chain.    

\section{Conclusions}
\label{sec-conclusion}
Motivated by recent experimental findings that show evidence for the 
competition or even coexistence of superconductivity and magnetism we 
have investigated the ground state properties of an itinerant $XY$ model. 
\begin{figure}[here]
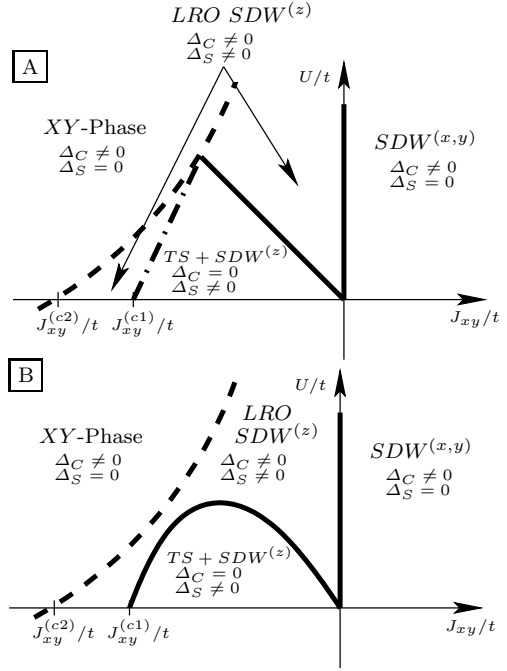

  \centering
  \input{weak_phase_diag_2.pstex_t}
  \input{weak_phase_diag_2b.pstex_t}
  \caption{\label{fig:dmrg_and_weak_phase}{The possible ground state 
      phase diagram of the itinerant $XY$-model at half-filling. Solid lines 
      mark second order phase transitions between the phases as obtained 
      from bosonization. The dashed line corresponds 
      to the spin-flop transition from the LRO SDW$^{(z)}$ into the 
      ferromagnetic $XY$ phase ($J_{xy}^{(c2)}\approx-4t$). 
      The dashed-dotted line marks the metal-insulator transition
      from the spin-gapped metallic phase with identical decay of triplet 
      superconducting and SDW$^{(z)}$ correlations into the LRO 
      antiferromagnetic (SDW$^{(z)}$) phase ($J_{xy}^{(c1)}\approx-3t$).
      Our numerical results do not exclude the possibility that the 
      multicritical point given in \small{$\fbox{A}$} is an artifact of the 
      bosonization approach. In a more realistic scenario of the phase 
      diagram the SDW$^{(z)}$ always separates the superconducting phase 
      from the easy-plane ferromagnetic phase as it is 
      shown \small{$\fbox{B}$}.}}
\end{figure}

First we considered the case of vanishing on-site Coulomb interactions.
The behavior of spin and charge gaps as function of the spin-coupling
$J_{xy}$ allows to distinguish four different phases (cf. with 
Fig.~\ref{fig:dmrg_and_weak_phase}). For antiferromagnetic
interactions $J_{xy} >0$ the spin gap vanishes, but the charge gap
is always finite. The observed behavior of the correlation functions
indicates a smooth evolution to the limiting case of spin-$1/2$ 
antiferromagnetic $XY$ chain at $J_{xy}\to \infty$.

For ferromagnetic couplings  $J_{xy} <0$ three different phases
appear. Already for weak interactions a spin gap opens, but the
charge sector is gapless. Here SDW$^{(z)}$ and triplet correlations, which 
decay with similar power-laws, are dominating, i.e.\ this regime exhibits
a coexistence of antiferromagnetic ordering and triplet superconductivity.
At $J_{xy}^{(c1)} \approx -3t$ the spin gap is maximal and a charge gap opens.
This intermediate phase, that extends up to $J_{xy}^{(c2)} \approx -4t$,
shows long-range order in the longitudinal spin correlation,
whereas superconducting correlations are suppressed and decay
exponentially as expected for the case of a finite charge gap.
Finally, at $J_{xy} > -4t$ via a spin-flop transition the system again 
enters a $XY$ phase characterized by vanishing spin but finite charge gap. 
Here the behavior is similar to the ferromagnetic $XY$ model.

The presence of a repulsive on-site Coulomb interaction $U$
has a strong effect on the phase diagram. Generically it leads
to an enlargement of the sectors with nonvanishing charge gap
at the expense of the sectors with spin gap. Already at $U=2t$
the charge gap is finite for all values of the exchange coupling $J_{xy}$.
Therefore the phase where antiferromagnetism and triplet
superconductivity coexist is no longer observed and magnetic
correlations become dominant everywhere. Only for small values
of the Coulomb interaction there is still a finite window
of coexistence possible.

Fig.~\ref{fig:dmrg_and_weak_phase} summarizes our findings. The phase 
diagram shown combines results from bosonization 
(cf. Fig.~\ref{fig:weak_phase}) and DMRG. However, it is difficult to 
determine the location of the phase boundaries numerically. E.g. it still 
has to be clarified whether the coexistence phase extends up to 
the $XY$-phase such that two LRO SDW$^{(z)}$ phases exist 
(Fig.~\ref{fig:dmrg_and_weak_phase}a). Alternatively, only one LRO 
SDW$^{(z)}$ phase exists such that it always separates the superconducting 
from the ferromagnetic $XY$-phase (Fig.~\ref{fig:dmrg_and_weak_phase}b).

\paragraph{Acknowledgments}

This work has been performed within the research program of
the SFB 608 funded by the DFG.
We like to thank Corinna Kollath and Ulrich Schollw\"ock for 
helpful information concerning the DMRG algorithm. We further thank
Erwin M\"uller-Hartmann and Achim Rosch for discussions.
GIJ also acknowledges support by the SCOPES grant N~7GEPJ62379

\end{document}